\documentclass[conference]{IEEEtran}
\usepackage{cite}

\ifCLASSINFOpdf
   \usepackage[pdftex]{graphicx}
\else
\fi
\usepackage{amsmath}
\usepackage{amssymb}
\usepackage{xcolor}
\usepackage[nolist,nohyperlinks]{acronym}

\usepackage{algorithm}
\usepackage[noend]{algpseudocode}
\usepackage{url}

\DeclareMathOperator*{\argmax}{argmax} 

\renewcommand{\Re}[1]{\mathfrak{Re}\left\{#1\right\}}
\renewcommand{\Im}[1]{\mathfrak{Im}\left\{#1\right\}}

\acrodef{UE}{user equipment}
\acrodef{BS}{base station}
\acrodef{Tx}{transmitter}
\acrodef{Rx}{receiver}
\acrodef{CFO}{carrier frequency offset}
\acrodef{ULA}{uniform linear array}
\acrodef{OFDM}{orthogonal frequency division multiplexing}
\acrodef{AWGN}{additive white Gaussian noise}
\acrodef{LO}{local oscillator}
\acrodef{MAP}{maximum \textit{a posteriori}}
\acrodef{mmWave}{millimeter wave}
\acrodef{subTHz}{sub-terahertz}
\acrodef{LOS}{line-of-sight}
\acrodef{NLOS}{non-line-of-sight}
\acrodef{MCMC}{Markov chain Monte Carlo}
\acrodef{VB}{variational Bayes}
\acrodef{EP}{expectation propagation}
\acrodef{KL}{Kullback-Leibler}
\acrodef{SBL}{sparse Bayesian learning}
\acrodef{RV}{random variable}
\acrodef{PDF}{probability density function}
\acrodef{EM}{expectation maximization}
\acrodef{5G}{fifth generation}
\acrodef{iid}{independent identically distributed}
\acrodef{ML}{maximum likelihood}
\acrodef{SCA}{successive convex approximation}
\acrodef{MM}{majorization-minimization}
\acrodef{ADMM}{alternating direction method of multipliers}
\acrodef{SNR}{signal-to-noise ratio}
\acrodef{AoA}{angle of arrival}
\acrodef{AoD}{angle of departure}
\acrodef{MSE}{mean squared error}
\acrodef{RMSE}{root mean squared error}
\acrodef{ReMSE}{relative mean squared error}
\acrodef{PSO}{particle swarm optimization}
\acrodef{NM}{Nelder-Mead}
\acrodef{B5G}{beyond 5G}
\acrodef{6G}{sixth generation standard}
\acrodef{MIMO}{multiple input multiple output}
\acrodef{LLN}{law of large numbers}
\acrodef{LGR}{large grid regime}
\acrodef{LPA}{long pilot approximation}
\acrodef{MC}{Monte Carlo}
\acrodef{ToF}{time of flight}
\acrodefplural{TOF}[TOFs]{times of flight}
\acrodef{CACC}{cross-antenna cross-correlation}
\acrodef{MMSE}{minimum mean square error}
\acrodef{DTFT}{discrete-time Fourier transform}
\acrodef{DFT}{discrete Fourier transform}
\acrodef{FFT}{Fast Fourier transform}
\acrodef{2P2T}{two precoders two transmissions}
\acrodef{TVP}{time-varying precoder}
\acrodef{RF}{radio frequency}
\acrodef{PSF}{point spread function}
\acrodef{DBSCAN}{density-based spatial clustering of applications with noise}
\acrodef{SSE}{Sum Square Error}
\acrodef{TDMA}{time division multiple access}
\acrodef{FDMA}{frequency division multiple access}
\acrodef{CDMA}{code division multiple access}
\acrodef{BER}{bit error rate}
\acrodef{MUSIC}{multiple signal classification}
\acrodef{URA}{uniform rectangular array}
\acrodef{JCS}{joint communication and sensing}
\acrodef{NPNT}{N-precoders N-transmissions}
\acrodef{SOR}{successive over-relaxation}
\acrodef{COMPAS}{concurrent mapping, positioning, and synchronization}
\acrodef{SLAM}{simultaneous localization and mapping}
\acrodef{AECD}{alternating exact coordinate descent}
\acrodef{ISAC}{integrated sensing and communication}
\acrodef{JSC}{joint sensing and communication}
\acrodef{FR2}{frequency range 2}
\acrodef{CSI}{channel state information}
\acrodef{SAGE}{space-alternating generalized expectation-maximization}

\usepackage{subcaption}
\usepackage{diagbox}

\begin{document}


%
\title{Sequential MAP Parametric OFDM Channel Estimation for Joint Sensing and Communication}
%
%
%

\author{\IEEEauthorblockN{Enrique~T.~R.~Pinto and Markku~Juntti}
\IEEEauthorblockA{Centre for Wireless Communications (CWC), University of Oulu, Finland\\
\{enrique.pinto, markku.juntti\}@oulu.fi}}

\markboth{Journal of \LaTeX\ Class Files,~Vol.~14, No.~8, August~2015}%
{Shell \MakeLowercase{\textit{et al.}}: Bare Demo of IEEEtran.cls for IEEE Journals}

%



\maketitle

\noindent
\begin{abstract}
Uplink sensing is still a relatively unexplored scenario in integrated sensing and communication which can be used to improve positioning and sensing estimates. We introduce a pilot-based maximum likelihood, and a maximum a posteriori parametric channel estimation procedure using an orthogonal frequency division multiplexing (OFDM) waveform in uplink sensing. The algorithm is capable of estimating the multipath components of the channel, such as the angles of arrival, departure, path coefficient, and the delay and Doppler terms. As an advantage, when compared to other existing methods, the proposed procedure presents expressions for exact alternating coordinate updates, which can be further improved to achieve a competitive multipath channel estimation tool.
\end{abstract}

\noindent

%
\IEEEpeerreviewmaketitle

%
%
%
%

\vspace{-0.3cm}
\section{Introduction}
\IEEEPARstart{R}{adio-based} sensing is being intensively studied for the purposes of \ac{JSC}. Exploiting the existing cellular infrastructure to perform sensing of passive devices, localization of active users, and mapping of the environment is not only economically attractive, it is also technically useful. Sensing, positioning, and environment data can not only be used to enhance mobile communications by improving power allocation, beamforming, and user scheduling, but it can also serve other systems such as autonomous vehicles and urban infrastructure by providing information for accident prevention, traffic flow optimization, etc.\par

As wireless communications standards progressively incorporate higher frequency ranges to their spectrum, such as \ac{FR2} in the \ac{5G} standard and also the very likely inclusion of \ac{subTHz} bands in \ac{B5G} and \ac{6G}, high mobility scenarios provide shorter and shorter channel coherence times. In these cases, \ac{CSI} acquisition becomes a non-trivial problem, as channel estimates quickly become outdated due to Doppler shifts, thus, only estimating the channel matrix stops being an effective option. Extracting geometrical propagation information and using it as a deterministic (or hybrid) channel model can be a useful method \cite{Tan}, especially because it paves the way for channel prediction and enviroment sensing/mapping. If the propagation parameters of each multipath are well estimated, the line between sensing with mapping and channel estimation becomes blurred; these values allow us to approximately reconstruct the channel with a deterministic model instead of consigning propagation phenomena to stochastic terms. Furthermore, they provide essential information for \ac{JSC}, which can be used to detect passive sensing targets, map the environment, and enhance the position estimates of users.\par

\vspace{-0.05cm}
In this paper, we propose a sequential \ac{MAP} parametric channel estimation method for extracting the parameters of each multipath component of the channel in the context of \emph{bistatic uplink sensing.}
The most popular solution in non-real-time channel modelling applications is the \ac{SAGE} algorithm \cite{sage }. While generally successful, its alternating coordinate descent often rely on line-search procedures. This can limit its applicability in real-time scenarios.
Other existing algorithms use the CANDECOMP/PARAFAC-decomposition (CP-decomposition) for a similar channel estimation procedure \cite{henk_cp, parafac_zhou}. However, they do not immediately exploit the structure of the channel tensor. Furthermore, the CP-decomposition is computationally expensive and outputs the best fitting rank $K$ decomposition of the input tensor, requiring further processing for extracting channel parameters.
In contrast, our proposed algorithm immediately outputs the channel parameters and exploits the channel model structure when computing their estimates, while also providing expressions for exact coordinate updates. This makes way for future work on improved channel estimation techniques that can further optimize the speed and accuracy of the channel parameter estimation process.\par

\vspace{-0.05cm}
The rest of the paper is structured as follows. In Section \ref{sec:model}, we introduce the model considered in this paper. Then, in Section \ref{sec:map_est}, we present the chosen estimation approach. In Section \ref{sec:opt_prelim}, we introduce the necessary background for the optimization algorithm that is proposed in Section \ref{sec:opt}. Finally, we analyse some numerical results in Section \ref{sec:numres} and make our concluding remarks in Section \ref{sec:conclusion}.

\vspace{-0.2cm}
\section{System Model}\label{sec:model}
Consider the following \ac{OFDM} uplink received signal model \cite{zhang_enabling_jsc}
\vspace{-0.3cm}
\begin{multline}
    \mathbf{y}_{n,t} = \sum^{L}_{\ell=1} b_{\ell} e^{-j2\pi n (\tau_{\ell}+\tau_{o})f_c} e^{j2\pi t (f_{D,\ell} + f_{o}) T_s} \\
    \cdot \mathbf{a}(\phi_{\ell}) \mathbf{a}^T(\theta_{\ell}) \mathbf{x}_{n,t} + \mathbf{w}_{n,t}, \label{eq:rec_sig}
\end{multline}
where $n$ and $t$ denote the \ac{OFDM} subcarrier and symbol index, respectively; $\mathbf{y}_{n,t}$ is the signal received by the \ac{BS} at the $n$th subcarrier and $t$th symbol; $L$ is the number of multipath components; $b_{\ell}$ is the $\ell$th path gain; $\tau_{\ell}$ is the propagation delay of the $\ell$th multipath; $\tau_{o}$ is the clock timing offset between the \ac{UE} and the \ac{BS}; $f_c$ is the subcarrier spacing $B/N_c$, where $B$ is the bandwidth; $f_{D,\ell}$ is the Doppler frequency of the $\ell$th multipath; $f_{o}$ is the \ac{CFO} of between \ac{UE} and the \ac{BS}; $T_s$ is the \ac{OFDM} symbol length; $\mathbf{a}(\phi/\theta)$ is the \ac{ULA} response vector with $N_r$/$N_t$ antennas and angle of arrival/departure $\phi/\theta$, given by $\mathbf{a}(\phi/\theta) = \begin{bmatrix} 1 & e^{-j\pi \sin(\phi/\theta)} & \cdots & e^{-j\pi(N_{R/T} -1) \sin(\phi/\theta)} \end{bmatrix}^T$, where ``$\phi/\theta$" here denotes ``either $\phi$ or $\theta$"; $\mathbf{x}_{n,t}$ is the transmitted pilot at the $n$th subcarrier and $t$th symbol; and finally $\mathbf{w}_{n}$ is \ac{AWGN} at the $n$th subcarrier and $t$th symbol with covariance $N_0 \mathbf{I}_{N_r}$. Because the signal is transmitted by the \ac{UE}, this scenario is called uplink sensing. Other variations of the uplink sensing also exist, those are based on setting up \acp{UE}, synchronized and with shared oscillator signals, deployed specifically for sensing. The model in (\ref{eq:rec_sig}) is general nonetheless, the dedicated \ac{UE} scenario is readily obtained by setting the offsets to zero.\par

In a communications context, we are usually exclusively interested in the composited values of the channel matrices
\vspace{-0.2cm}
\begin{equation}
    \mathbf{H}_{n,t} 
    =  \sum^{L}_{\ell=1} b_{\ell} e^{j \omega_{1,\ell} n } e^{j \omega_{2,\ell} t} \mathbf{a}(\phi_{\ell}) \mathbf{a}^T(\theta_{\ell}), \label{eq:H}
\end{equation}
where $\omega_{1,\ell} =-2\pi(\tau_{\ell}+\tau_{o})f_c$ and $\omega_{2,\ell} = 2\pi(f_{D,\ell} + f_{o}) T_s$.
However, in radio-based sensing and localization we are interested in estimating the sensing parameters $(b, \tau, f_{D}, \phi, \theta)$. Furthermore, the timing and frequency offset parameters $\boldsymbol{\xi}_o = (\tau_{o}, f_{o})$ are important to be estimated, because they lead to ranging and speed estimation ambiguity. It may be assumed that the offsets are the same for all the antennas, because the signal from the \ac{LO} is shared within the radio chains of an \ac{UE}. In this work, we do not tackle the estimation of the offsets, instead we focus exclusively on estimating $\boldsymbol{\xi}_\ell = (b_\ell, \omega_{1,\ell}, \omega_{2,\ell}, \phi_\ell, \theta_\ell) \forall \ell$. The offsets remain a nuissance and additional estimation methods would be required to identify them if the times-of-flight or Doppler frequencies are of interest.

\vspace{-0.15cm}
\section{Maximum a Posteriori Estimation} \label{sec:map_est}
Define $\mathbf{y}=\text{vect}(y_{n,t,u})$, where $\text{vect}(\cdot)$ denotes the tensor vectorization operation, also denote by $\boldsymbol{\xi}$ the vector of sensing parameters $\boldsymbol{\xi}_\ell$ for all detected paths, then the posterior of $\boldsymbol{\xi}$ given the data $\mathbf{y}$ is
\begin{equation}
    p(\boldsymbol{\xi}|\mathbf{y}) = \frac{p(\mathbf{y}|\boldsymbol{\xi}) p(\boldsymbol{\xi})}{p(\mathbf{y})} = \frac{\prod_{n,t,u} p(y_{n,t,u}|\boldsymbol{\xi}) p(\boldsymbol{\xi})}{\int_\Xi \prod_{n,t,u} p(y_{n,t,u}|\boldsymbol{\xi}')p(\boldsymbol{\xi}')  d\boldsymbol{\xi}'}, \label{eq:posterior_xi}
\end{equation}
where $\Xi$ denotes the parameter space and $p(\boldsymbol{\xi})$ denotes the prior for $\boldsymbol{\xi}$. Throughout the remainder of this paper, summation and products over $n/t/u/v$ go from 0 to $N_{c/s/r/t}-1$, unless otherwise indicated. The \ac{MAP} estimate is then given by
\begin{equation}
    \hat{\boldsymbol{\xi}} = \argmax_{\boldsymbol{\xi}} p(\boldsymbol{\xi}|\mathbf{y}) = \argmax_{\boldsymbol{\xi}} \prod_{n,t,u} p(y_{n,t,u}|\boldsymbol{\xi}) p(\boldsymbol{\xi}),
\end{equation}
since the denominator of (\ref{eq:posterior_xi}) is a constant. The conditional \ac{PDF} of the data is complex normal $ y_{n,t,u}|\boldsymbol{\xi} \sim \mathcal{CN}\left(\mu_{n,t,u}(\boldsymbol{\xi}) ,N_0 \right)$, where the mean is given by 
\vspace{-0.3cm}
\begin{equation}
    \mu_{n,t,u}(\boldsymbol{\xi}) = \sum^{L}_{\ell=1} b_{\ell} e^{j \omega_{1,\ell} n } e^{j \omega_{2,\ell} t} e^{-j\pi u \sin(\phi_\ell)} \mathbf{a}^T(\theta_{\ell})\mathbf{x}_{n,t}. \label{eq:mean}
\end{equation}
The priors are assumed to be independent.
Given the likelihood and prior, the log-posterior is
\begin{multline}
    \log p(\boldsymbol{\xi}|\mathbf{Y}) = -\frac{1}{N_0}\sum_{n,t,u} \left| y_{n,t,u} - \mu_{n,t,u}(\boldsymbol{\xi}) \right|^2 \\
    + \sum^{L}_{\ell=1}  \log (p(b_\ell)p(\omega_{1,\ell})p(\omega_{2,\ell})p(\phi_{\ell})p(\theta_{\ell}))  + \dots,
\end{multline}
where we have omitted the constant terms. \par 

\vspace{-0.1cm}
\section{Optimization Preliminaries}\label{sec:opt_prelim}
We write the \ac{MAP} estimation as a constrained minimization problem
\begin{align}
       \min_{\boldsymbol{\xi}} &\left[ \frac{1}{N_0}\sum_{n,t,u} \left| y_{n,t,u} - \mu_{n,t,u}(\boldsymbol{\xi}) \right|^2 - \log p(\boldsymbol{\xi}) \right]  \label{eq:opt_obj}\\
       \text{s.t. }\quad &\angle b_\ell,\, \omega_{1,\ell},\, \omega_{2,\ell} \in (-\pi.\pi);\; \phi_\ell,\, \theta_\ell \in \left( -\frac{\pi}{2}, \frac{\pi}{2} \right) \; \forall \ell. \label{eq:opt_constr}
\end{align}
The objective function is clearly nonconvex over $\boldsymbol{\xi}$ and is $5L$-dimensional, which can be quite high if there are many multipaths. For this reason, simple local descent methods, such as gradient descent and its variations, are not effective. 
Additionally, the objective function computation can be quite expensive if the number of receive antennas, subcarriers, and \ac{OFDM} symbols is large. The computational cost for objective function evaluation makes many global optimization methods, such as particle swarm and simulated annealing, extremely time consuming until an acceptable solution is achieved. One technique that is successful for this problem is an augmented form of \ac{AECD}, further details are provided in Section \ref{sec:opt}.\par

To perform exact coordinate descent we require that the gradient along that coordinate direction be equal to zero, e.g. for the angle of arrival of path $\ell'$ we have $\frac{\partial f}{\partial \phi_\ell'} = 0$, where $f$ denotes the objective function in (\ref{eq:opt_obj}). Breaking down the objective function into the sum of the log-likelihood and the log-prior terms, respectively, we have $f(\boldsymbol{\xi},\mathbf{y})=\log p(\mathbf{y}|\boldsymbol{\xi})+\log p(\boldsymbol{\xi})$. We will show that the partial derivatives of the log-likelihood term with relation to $\phi_{\ell'}$, $\theta_{\ell'}$, $\omega_{1,\ell'}$, and $\omega_{2,\ell'}$, are given by Fourier series. The series has as many terms as the size of that parameters associated dimension, e.g. $\frac{\partial \log p(\mathbf{y}|\boldsymbol{\xi})}{\partial \phi_\ell'}$ has $N_r$ terms, $\frac{\partial \log p(\mathbf{y}|\boldsymbol{\xi})}{\partial \omega_{1,\ell'}}$ has $N_c$ terms, and so on. The roots of the resulting series (including the additional prior term) will be candidate solutions for the coordinate descent update. The Fourier series root-finding problem can be turned into a companion matrix eigenvalue problem \cite{fourier_roots}, we can thus readily find all roots by applying a transformation to the computed eigenvalues. Finally, we evaluate the objective on all the roots and select the one with smallest value.\par
We now present the partial derivatives of $\log p(\mathbf{y}|\boldsymbol{\xi})$ over the $\phi_{\ell'}$, $\theta_{\ell'}$, $\omega_{1,\ell'}$, and $\omega_{2,\ell'}$ coordinates. We omit the derivation for space constraints. Over the following section, some indices will be arbitrarily moved from subscript to superscript in order to save space. Additionally we denote the transmitted signal at transmit antenna $v$ as $x^v_{n,t}$.

\vspace{-0.2cm}
\subsection{Partial Derivative over $\omega_{1,\ell'}$ and $\omega_{2,\ell'}$}
The partial derivative over $\omega_{1,\ell'}$ is given by
\begin{gather}
    \frac{\partial \log p(\mathbf{y}|\boldsymbol{\xi})}{\partial \omega_{1,\ell'}} = \sum^{N_c-1}_{n=0} a_n\cos(\omega_{1,\ell'} n) + b_n \sin(\omega_{1,\ell'} n) \label{eq:omega1_fourier} \\
     a_n = \frac{2n}{N_0} \sum_{t,u} \Im{\alpha^u_{\ell',n,t}\left( y^{u,*}_{n,t} -  \sum_{\ell\neq\ell'} e^{-j\omega_{1,\ell} n} \alpha^{u,*}_{\ell,n,t} \right)}\\
     b_n = \frac{2n}{N_0} \sum_{t,u} \Re{\alpha^u_{\ell',n,t}\left( y^{u,*}_{n,t} - \sum_{\ell\neq\ell'} e^{-j\omega_{1,\ell} n} \alpha^{u,*}_{\ell,n,t}  \right)} \\
     \alpha^u_{\ell,n,t} = b_\ell e^{j\omega_{2,\ell}t} e^{-j\pi u\sin(\phi_\ell)} \mathbf{a}^T(\theta_\ell) \mathbf{x}_{n,t}.
\end{gather}
The partial derivative over $\omega_{2,\ell'}$ is similar, by symmetry.

\subsection{Partial Derivative over $\sin(\phi_{\ell'})$}
For $\phi_{\ell'}$, we take the derivative over $\sin(\phi_{\ell'})$ and exploit the bijectivity of the sine function over the $(-\frac{\pi}{2},\frac{\pi}{2})$ range to compute the value of $\phi_{\ell'}$ that satisfies $\frac{\partial \log p(\mathbf{y}|\boldsymbol{\xi})}{\partial\sin(\phi_{\ell'})}=0$ with smallest objective value. The partial derivative is given by
\begin{gather}
     \frac{\partial \log p(\mathbf{y}|\boldsymbol{\xi})}{\partial\sin(\phi_{\ell'})} = \hspace{-6pt} \sum^{N_R-1}_{u=0} a_u\cos(\pi u\sin(\phi_{\ell'})) + b_u \sin(\pi u\sin(\phi_{\ell'})) \label{eq:sinphi_fourier} \\
     a_u = \hspace{-1pt} \frac{2u}{N_0} \sum_{t,n} \Im{\alpha^*_{\ell',n,t}\left( y^{u}_{n,t} - \sum_{\ell\neq\ell'} e^{-j\pi u\sin(\phi_\ell)} \alpha_{\ell,n,t} \right)}\\
     b_u = \hspace{-1pt} \frac{2u}{N_0} \sum_{t,n} \Re{\alpha^*_{\ell',n,t}\left( y^{u}_{n,t} -  \sum_{\ell\neq\ell'} e^{-j\pi u\sin(\phi_\ell)} \alpha_{\ell,n,t}  \right)}\\
     \alpha_{\ell,n,t} = b_\ell e^{j\omega_{1,\ell}n} e^{j\omega_{2,\ell}t}  \mathbf{a}^T(\theta_\ell) \mathbf{x}_{n,t}.
\end{gather}

\subsection{Partial Derivative over $\sin(\theta_{\ell'})$}
Once again, exploiting the injectivity of the sine function, we get
\begin{equation}
    \frac{\partial \log p(\mathbf{y}|\boldsymbol{\xi})}{\partial\sin(\theta_{\ell'})} = \sum^{N_T-1}_{v=0} a_v \cos(\pi v\sin(\theta_{\ell'})) + b_v \sin(\pi v\sin(\theta_{\ell'})), \label{eq:sintheta_fourier}
\end{equation}
where the coefficients are given by $a_v = \frac{2}{N_0}\sum_{n,t,u} v \alpha_{n,t,u,v}$ and $b_v = -\frac{2}{N_0}\sum_{n,t,u} v \beta_{n,t,u,v}$,
which in turn are expressed in terms of $\alpha_{n,t,u,v}$, given by
\begin{multline}
     \alpha_{n,t,u,v} =\Re{\gamma^{u}_{\ell',n,t}}\Im{x^v_{n,t}} + \Im{\gamma^{u}_{\ell',n,t}}\Re{x^v_{n,t}} \\  + |\alpha^u_{\ell',n,t}|^2\Im{\sum^{N_T-1}_{k=v}  x^{k}_{n,t}x^{{k-v},*}_{n,t}} -\Im{y^{u,*}_{n,t} \alpha^u_{\ell',n,t}x^v_{n,t}}; 
\end{multline}
\vspace{-0.1cm}
and $\beta_{n,t,u,v}$. For $v=0$:
\begin{multline}
     \beta_{n,t,u,0} =  \Re{\gamma^{u}_{\ell',n,t}}\Re{x^v_{n,t}} - \Im{\gamma^{u}_{\ell',n,t}}\Im{x^v_{n,t}} \\
        +\frac{|\alpha^u_{\ell',n,t}|^2}{2} \sum^{N_T-1}_{k=0} |x^k_{n,t}|^2 -\Re{y^{u,*}_{n,t} \alpha^u_{\ell',n,t}x^v_{n,t}}
\end{multline}
\vspace{-0.1cm}
and, for $v=1,\dots,N_t-1$:
\begin{multline}
     \beta_{n,t,u,v} = \Re{\gamma^{u}_{\ell',n,t}}\Re{x^v_{n,t}} - \Im{\gamma^{u}_{\ell',n,t}}\Im{x^v_{n,t}}\\
        + |\alpha^u_{\ell',n,t}|^2\Re{\sum^{N_T-1}_{k=v}  x^{k}_{n,t}x^{{k-v},*}_{n,t}} -\Im{y^{u,*}_{n,t} \alpha^u_{\ell',n,t}x^v_{n,t}}.
\end{multline}

\vspace{-0.5cm}
\subsection{Optimization over $b_{\ell'}$} \label{subsec:b_update}
We assume a complex normal prior for $b_{\ell'}$, with mean $\Bar{b}_{\ell'}$ and variance $\nu_{b_{\ell'}}$. It can be seen that $f$ is convex over $b_{\ell'}$. Using Wirtinger calculus, we can derive closed form expressions for the exact coordinate update on $b_{\ell'}$, for a single $\ell'$ (even though a closed form joint update for $b_\ell \forall \ell$ exists by solving a linear system). We once again omit the derivation, presenting only the result
\begin{equation}
    b^{\text{opt}}_{\ell'} = \frac{ \nu_{b_{\ell'}}\sum_{n,t,u} \gamma^{u,*}_{\ell',n,t} \left( y^u_{n,t} - \sum_{\ell\neq\ell'} b_\ell\gamma^u_{\ell,n,t} \right) + N_0 \Bar{b}_{\ell'}}{\nu_{b_{\ell'}}\sum_{n,t,u} |\gamma^u_{\ell',n,t}|^2 + N_0}, \label{eq:b_opt}
\end{equation}
where $\gamma_{\ell,n,t,u} = e^{j\omega_{1,\ell}n} e^{j\omega_{2,\ell}t} e^{-j\pi u \sin(\phi_\ell)} \mathbf{a}^T(\theta_\ell) \mathbf{x}_{n,t}$.

\vspace{-0.1cm}
\subsection{Priors}
Because we want to preserve the Fourier series structure of the partial derivatives, we must choose priors which have derivatives that can be directly incorporated into a Fourier series.
For $\omega_{1.\ell}$, we consider the following prior distribution
\begin{equation}
    p(\omega_{1.\ell}) \propto \exp \left( -\frac{|e^{j\Bar{\omega}_{1,\ell}n} - e^{j \omega_{1,\ell} n}|^2}{\nu_{\omega_{1,\ell}}}  \right),
\end{equation}
where $\Bar{\omega}_{1,\ell}\in(-\pi,\pi)$ and $\nu_{\omega_{1,\ell}}>0$ respectively denote the mode and variance parameter. Note that, while the mode of the distribution is indeed equal to $\Bar{\omega}_{1,\ell}$, the variance is merely an increasing function of $\nu_{\omega_{1,\ell}}$. A similar prior is used for $\omega_{2.\ell}$. For $\phi_{\ell}$ we use
\begin{equation}
    p(\phi_{\ell}) \propto \exp \left( -\frac{|e^{j\pi\sin(\Bar{\phi}_{\ell})} - e^{j\pi\sin(\phi_{\ell})}|^2}{\nu_{\phi_{\ell}}}  \right),
\end{equation}
with mode and variance parameters similarly defined. The proposed prior for $\theta_\ell$ is identical. The path gain coefficient prior has been already introduced in Subsection \ref{subsec:b_update}. In sequential estimation, the mode of the current estimation step corresponds to the point estimates of the previous step, the variance however must be heuristically chosen.

\vspace{-0.2cm}
\subsection{Partial Derivative of the Priors}
The presented partial derivatives include only the log-likelihood term. We must add the log-prior to have the complete objective. The derivative of the log-prior of $\phi_{\ell'}$ is
\begin{multline}
    \frac{\partial \log p(\boldsymbol{\xi})}{\partial \sin(\phi_{\ell'})} = -\frac{2\pi}{\nu_{\phi_{\ell'}}}\sin(\pi \sin(\Bar{\phi}_{\ell'}))\cos(\pi\sin(\phi_{\ell'}))\\
    + \frac{2\pi}{\nu_{\phi_{\ell'}}}\cos(\pi \sin(\Bar{\phi}_{\ell'})\sin(\pi\sin(\phi_{\ell'})).
\end{multline}
A similar equation applies for $\theta_{\ell'}$, by symmetry. For $\omega_{1,\ell'}$ we have
\begin{equation}
    \frac{\partial \log p(\boldsymbol{\xi})}{\partial \omega_{1,\ell'}} \! = \! \frac{2\cos(\Bar{\omega}_{1,\ell'})\sin(\omega_{1,\ell'})}{\nu_{\omega_{1,\ell'}}} - \frac{2\sin(\Bar{\omega}_{1,\ell'})\cos(\omega_{1,\ell'})}{\nu_{\omega_{1,\ell'}}} .
\end{equation}
Again, the expression for $\omega_{2,\ell'}$ follows by symmetry. By adding these terms to the partial derivatives of the log-likelihood term we get the partial derivative of the objective.
\vspace{-0.6cm}
\section{Optimization Procedure} \label{sec:opt}
For the inference problem above, the gradient or coordinate descent methods by themselves are ineffective in providing acceptable solutions. Also, due to the dimensions and evident nonconvexity of the optimization problem, proving optimality of the solutions is hard. To achieve a useful feasible solution, we propose an \ac{AECD} method, in which the parameters for a single multipath index are optimized in an exact alternating fashion in an inner loop, while the outer loop varies the current multipath index. Because the exact coordinate descent is still a local descent method, we augment it with a combination of momentum and a \ac{SOR} inspired coordinate update, this is essential to escape local optima and improve the estimation results.\par
Let us first detail the outer and inner loop structure. First, a maximum number of expected paths $L_{\text{max}}$ is defined. This number should be surely larger than the possible number of detectable paths, i.e., paths with power that is not much smaller than the noise variance, and depends heavily on the propagation characteristics of the environment. The outer loop progresses along path indices in the following order: 
\begin{equation}
    \mathbf{I} = \left[1, 2, 1, 2, 3, 1, 2, 3, 4, \dots, L_{\text{max}}-1, L_{\text{max}},1,\dots,L_{\text{max}} \right]. \label{eq:path_order}
\end{equation}
Intuitively, after the first path is detected and roughly estimated, the algorithm moves on to detect the next path. Once the next path is detected and estimated, then the algorithm returns to the first path such as to ``compensate the interference" of the previously undetected second path when estimating the first path. This reasoning proceeds until hopefully all paths up to $L_{\text{max}}$ have been estimated. If at some point of the outer loop no more paths remain, then the algorithm starts outputing spurious paths, which have no physical correspondence. This means that choosing a large value for $L_{\text{max}}$ has a time cost, as the algorithm would have to estimate many spurious paths before finishing. It is convenient to devise a procedure to detect when all true paths have already been detected.\par

\begin{algorithm}[!b]
\caption{Multipath parameter estimation algorithm.} 
\label{alg:algo1}
\begin{algorithmic}[1]
    \Procedure{EstimateParams}{$\mathbf{y}$, $L_{\text{max}}$, $\hat{\boldsymbol{\xi}}_{i-1}$}       
    \State $l=1$;
    \For{$\ell = \mathbf{I}[l]$} \Comment{Path order $\mathbf{I}[l]$ as in defined (\ref{eq:path_order}})
        \State Initialize $\boldsymbol{\xi}_{\ell}=\mathbf{0}$;
        \For{$\text{it}=1,\dots,\text{it}_\text{max}$}
            \State Compute objective $f_{0} = f(\boldsymbol{\xi},\mathbf{y})$;
            \State Update coordinates in the order: $b_{\ell}$, $\omega_{1,\ell}$, $b_{\ell}$, $\omega_{2,\ell}$, $b_{\ell}$, $\theta_{\ell}$, $b_{\ell}$, using (\ref{eq:var_update1}) and (\ref{eq:var_update2});
            \State Compute objective $f_{1} = f(\boldsymbol{\xi},\mathbf{y})$;
            \If{$|\boldsymbol{\Delta}(\boldsymbol{\xi}_{\ell})|\prec\epsilon_{\text{var}}$}
                \State Break;
            \ElsIf{$f_1-f_0 < \epsilon_{\text{obj}}$}
                \State Break;
            \ElsIf{Other stopping heuristics}
                \State Break;
            \EndIf
        \EndFor
        \State $l=l+1$;
    \EndFor
    \State Estimate $\hat{L}$; \label{state:estimate_L}
    \State \Return $\boldsymbol{\xi}$; 
\EndProcedure
\end{algorithmic}
\end{algorithm}

\begin{algorithm}[!b]
\caption{Algorithm for the estimation of the number of active paths.}
\label{alg:algo2}
\begin{algorithmic}[1]
    \Procedure{Estimate\_L}{$\mathbf{y}$, $\hat{\boldsymbol{\xi}}_{i}$, $\epsilon_L$}       
    \State Sort $\hat{\boldsymbol{\xi}}_\ell$ in decreasing $|b_\ell|$ order, for $\ell=1,\dots,L_{\text{max}}$;
    \State Save sorting order in vector $\mathbf{S} = [s_1,\dots,s_{L_{\text{max}}}]$
    \For{$i = 1,\dots,L_{\text{max}}$}
        \State Compute objective $f_i = f(\boldsymbol{\xi}_{s_1},\dots,\boldsymbol{\xi}_{s_i},\mathbf{y})$
    \EndFor
    \For{$i = 1,\dots,L_{\text{max}}$}
        \If{$f_i - f_{i+1} < \epsilon_L \frac{(f_1 - f_{i})}{i}$ }
        \State \Return $i$; 
        \EndIf
    \EndFor
    \State \Return $L_{\text{max}}$; 
\EndProcedure
\end{algorithmic}
\end{algorithm}

Moving on to the inner loop. Suppose that the current path at the outer loop is $\ell'$, then, in a single iteration, the coordinates are updated in the following order: $b_{\ell'}$, $\omega_{1,\ell'}$, $b_{\ell'}$, $\omega_{2,\ell'}$, $b_{\ell'}$, $\theta_{\ell'}$, $b_{\ell'}$, $\phi_{\ell'}$. The inner loop is repeated for a maximum set amount of iterations $\text{it}_\text{max}$. Updating the path coefficient $b_{\ell'}$ in-between the other coordinates apparently provides more efficient updates. Exploring this idea, for future work, it may be effective to define a ``new" objective function by direct substitution of the optimal paths using (\ref{eq:b_opt}) on (\ref{eq:mean}), and then attempt to optimize this function.\par
Finally, we describe the individual coordinate updates. Denote by $\xi_m$ the coordinate to be updated for the $m$th time, also denote by $\xi^{\text{opt}}_m$ its optimal coordinate descent update. Then its partial update with momentum is
\begin{equation}
    \xi'_{m+1} = \xi^{\text{opt}}_m + \eta_m (\xi_m - \xi_{m-1}), \label{eq:var_update1}
\end{equation}
where $\eta_m$ is the momentum coefficient of that variable at the $m$th update. We then perform a \ac{SOR} inspired rule to complete the coordinate update 
\vspace{-0.2cm}
\begin{equation}
    \xi_{m+1} = \text{Wrap}_\xi\left( (1-\lambda_m) \xi_m + \lambda_m \xi'_{m+1} \right), \label{eq:var_update2}
\end{equation}
where $\lambda_m\in[0.5,1.5]$ is the \ac{SOR} coefficient of that variable at update $m$, and $\text{Wrap}_\xi(\cdot)$ denotes wrapping the argument value to the valid domain of the parameter, e.g., $\phi$ and $\theta$ should be wrapped to the interval $(-\frac{\pi}{2},\frac{\pi}{2})$ and $\omega_1$ and $\omega_2$ to $(-\pi,\pi)$. Because each variable is updated with forward substitution (like a Gauss-Seidel update for solving linear equations), instead of updating all coordinates together (like a Jacobi update), we apply a heuristic form of \ac{SOR}, which is known to outperform the Gauss-Seidel for linear equations. While there are no theoretical convergence speed guarantees, it provides an additional degree of freedom to tune the algorithm.\par

An outer loop iteration may be interrupted and skipped if the estimates have failed to change by the desired amount in an inner loop iteration, e.g., if all parameters have not changed by more than $10^{-5}$. An outer loop iteration may also be skipped if the objective function has not changed by more than a threshold for a particular inner iteration. Given a set of multipath parameters from a previous estimation $\hat{\boldsymbol{\xi}}_{i-1}$, a basic outline of the proposed algorithm is provided in Algorithm \ref{alg:algo1}. In Algorithm \ref{alg:algo1}, $\boldsymbol{\xi}_{\ell}$ denotes the variables associated with path $\ell$, $\boldsymbol{\Delta}(\boldsymbol{\xi}_{\ell})$ denotes the vector of relative changes of all variables from path $\ell$, the inequality $|\boldsymbol{\Delta}(\boldsymbol{\xi}_{\ell})|\prec\epsilon_{\text{var}}$ denotes that all relative changes are less than the threshold $\epsilon_{\text{var}}$. Similarly, $\epsilon_{\text{obj}}$ is the threshold for objective change in a single iteration. One may consider using additional stopping heuristics such as keeping track of a trailing moving average, if some property of the moving average indicates slow convergence, then break and move on to the next outer loop iteration. \par

Line \ref{state:estimate_L} of Algorithm \ref{alg:algo1} requires estimating the number of paths. For this, we propose a method based on objective function decrease. It consists first sorting paths in decreasing order based on the estimated path powers $|b_\ell|$, then progress through the vector by including more paths, computing the objective function, and checking how much the objective decreased by including the last path. Proceed until a (possibly variable) threshold value is reached. The version used in Section \ref{sec:numres} is displayed in Algorithm \ref{alg:algo2}.

\vspace{-0.1cm}
\section{Numerical Results}\label{sec:numres}
\begin{figure}[!b]
\centering
\begin{subfigure}{.5\linewidth}
  \centering
  \includegraphics[width=0.7\textwidth]{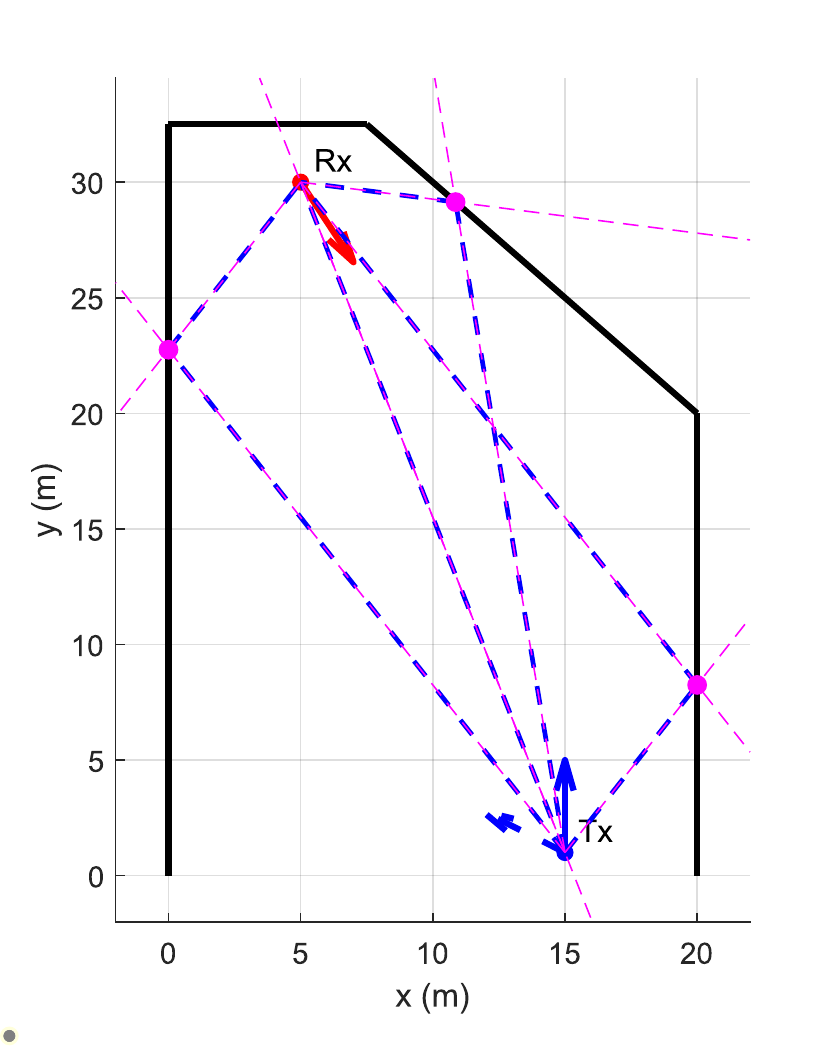}
  \caption{Single point.}
  \label{fig:sub1}
\end{subfigure}%
\begin{subfigure}{.5\linewidth}
  \centering
  \includegraphics[width=0.7\textwidth]{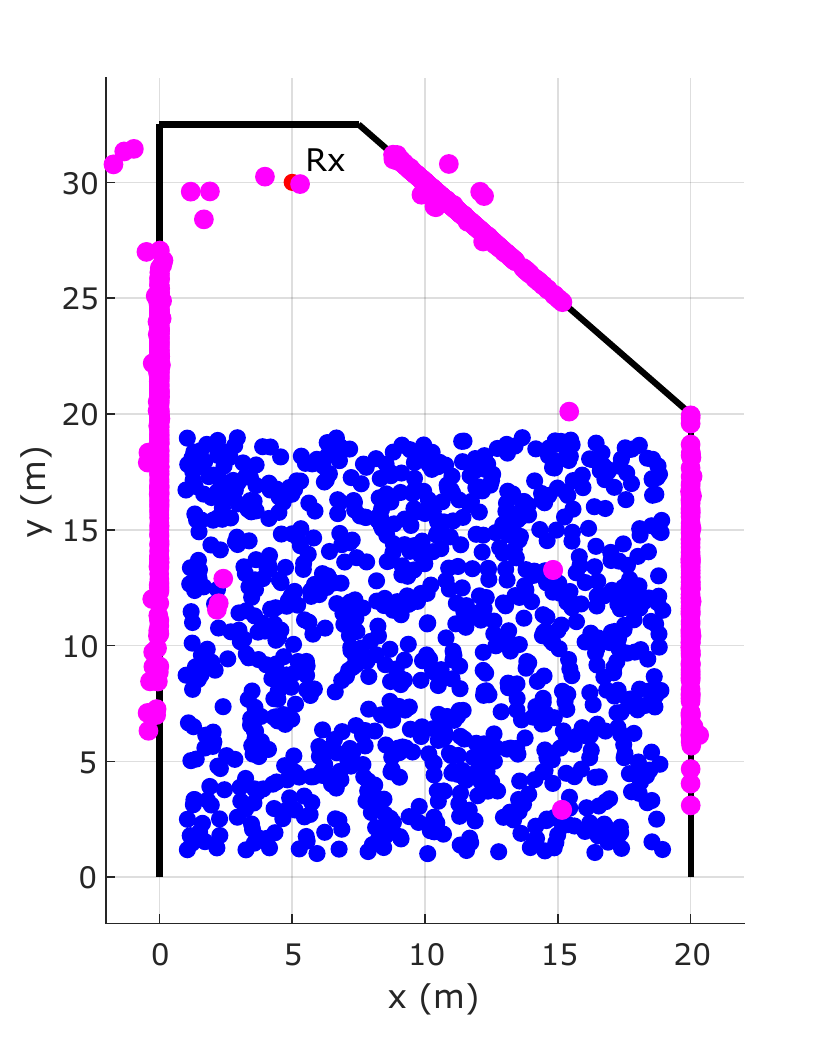}
  \caption{Ensemble.}
  \label{fig:sub2}
\end{subfigure}
\caption{Geometric channel simulation environment. The walls are represented by black lines, the propagation paths are indicated by blue dashed lines, the magenta dashed lines are the propagation directions of the signal computed from the \ac{AoA} and \ac{AoD} estimates (given perfect \ac{UE} and \ac{BS} pose information). The red/blue dot and arrow indicate the position and orientation of the \ac{BS}/\ac{UE}, respectively. The magenta dots are the intersection of the magenta lines, this is one way of estimating the reflector positions. The dashed blue line shows the transmitter velocity.}
\label{fig:environment_sim}
\end{figure}

In this section, we will assess the performance of the proposed method by analysing simulation results. Initially, we want to verify how effectively the algorithm detects the existing paths without any prior information. Then, we present a simple example of how this algorithm can be used for mapping, given perfectly known positions and orientations (poses) of the transmitter (\ac{UE}) and receiver (\ac{BS}). In both scenarios, we consider a transmtted pilot signal with 50 \ac{OFDM} symbols and 40 subcarriers. The transmitter and receiver have \acp{ULA} with 4 and 16 antennas, respectively. The used carrier frequency is 60~GHz, the subcarrier spacing is 240~kHz, and the symbol time is 4.46~$\mu$s, which corresponds to numerology $\mu=4$ in the \ac{5G} standard. The channel simulation considers only first order specular reflections. Paths with angles of arrival or departure outside the $\left(-\frac{\pi}{2},\frac{\pi}{2}\right)$ interval are considered to have zero gain. The environment used in this section is depicted in Fig.~\ref{fig:environment_sim}. For space constraints, we leave a detailed comparison with other methods \cite{sage, henk_cp, parafac_zhou} for future work.\par



We introduce a channel model with the intention of offerring a sufficient geometrical representation of multipath propagation for our estimation problems. The path coefficient is computed from the total propagation distance $d^2_\ell$ with an added power reflection loss $0<c_\ell<1$ if the path is not \ac{LOS}, given by $|b_\ell| = \sqrt{c_\ell/(4\pi d^2_\ell)}$.
We consider the transmit power $P_T$ to be equally allocated to all subcarriers $N_c$. Naturally, if the path is \ac{LOS}, then $c_\ell=1$. The reflection coefficient for \ac{NLOS} paths is set to $c_\ell=0.2$. The phase is sampled from a uniform distribution $\angle b_\ell \sim \mathcal{U}\left(-\pi,\pi\right)$, thus $b_\ell = |b_\ell|e^{j\angle b_\ell}$. The \acsp{ToF} are simply the path distance divided by the speed of light $\tau_\ell = d_\ell/c$. The Doppler frequency is computed from the projection of the \ac{UE} velocity on the departure direction vector $v_\ell$, and is given by $f_{D,\ell} = f_{\text{carrier}} v_\ell/c$. We consider a $\tau_o = 0.1~\mu$s clock offset between \ac{UE} and \ac{BS}. The carrier frequency offset is set to 2.4~MHz, 40~ppm of the carrier frequency.\par

For the model to be identifiable, the transmitted signal cannot be arbitrarily chosen. Intuitively, \ac{AoD} estimation requires that different angles of departure produce distinguishable outputs throughout the pilot sequence. It is impossible to estimate $\theta_\ell$ if a single data stream is transmitted with a fixed precoder. Using more data streams is one way to ensure that it is possible to estimate the \ac{AoD}. In the uplink context, it is not usual to transmit many streams. By transmitting a single stream, but varying the precoder, it is possible to guarantee identifiability. We consider 1 data stream and a time-varying precoder matched to angle $\Bar{\theta}\in\left(-\frac{\pi}{2},\frac{\pi}{2}\right)$, which is uniformly swept from $-\frac{\pi}{2}$ to $\frac{\pi}{2}$ during the 50 \ac{OFDM} symbols. \par

Given no prior, we want to assess the precision and recall of the path detection and estimation. We simulate 1024 different scenarios, with random \ac{UE} poses and \ac{BS} at $(5,30)$. The transmitter positions are uniformly distributed on the $[1,19]\times[1,19]$ rectangle, while their orientation is uniformly distributed on the $\psi_{\text{UE}\to\text{BS}}+[-\pi/2,\pi/2]$ interval, where $\psi_{\text{UE}\to\text{BS}}$ is the orientation where the \ac{UE} perfectly faces the \ac{BS}. Transmit power is set to 8~W, i.e., 9~dBW, equally divided along all subcarriers so that each subcarrier has $-7$~dBW. Noise power is set to $-80$~dBW. 
For the optimizer parameters, we set $L_{\text{max}}=6$, and the thresholds to $\epsilon_{\text{var}} = 10^{-5}$ and $\epsilon_{\text{obj}} = 10^{-6}$. The momentum coefficients are initialized to 0.1 and are decremented at every inner loop iteration with the rule $\eta_{m+1} = 0.99\eta_m$. The SOR coefficients are updated at every inner loop iteration with the following rule $\lambda_{\text{it}} = 0.98 + 0.22 \exp\left(-\frac{\text{it}}{15}\right)$, where ``$\text{it}$" is the inner loop iteration counter. The threshold for Algorithm \ref{alg:algo2} is $\epsilon_L = 0.5$. An example of the estimation results in this setup is represented by the magenta lines in Fig.~\ref{fig:sub1}. The full ensemble of points is shown in Fig.~\ref{fig:sub2}.\par 

It may happen that some paths do not converge but are still declared to be valid paths by our algorithm. To determine how frequently this happens, we compare the estimated paths to the true paths by computing $\|\boldsymbol{\xi}_\ell - \hat{\boldsymbol{\xi}}_\ell\|_2$ and performing greedy assignment. The estimated paths that had no assigned true paths were considered misdetections. On the other hand, estimated paths that were properly assigned to a true path were considered true detections. This way it is possible to estimate the Precision and Recall of our algorithm. Following the described procedure yields precision of 0.9938 and recall of 0.9854. The estimates of true detections have their \ac{MSE} and \ac{RMSE} values shown in Table \ref{tab:RMSE_ml}. It can be seen that, ignoring misdetections, the quality of estimates is quite useful, particularly for the $\omega_1$ and $\omega_2$ values as well as the path magnitude $|b_\ell|$. The estimates for the angles of departure and arrival are not as good, but are still sufficient for approximately sensing the environment, given perfect transmitter pose information.
Using $\omega_1$ and $\omega_2$ requires very fine clock and carrier synchronization to eliminate the offsets and extract useful geometric information. \par

\vspace{-0.2cm}
\begin{table}[!htbp]
\caption{\ac{MSE} and \ac{RMSE} for the parameter estimates of valid paths, \ac{ML} estimation}
\begin{center}
\resizebox{\columnwidth}{!}{
\begin{tabular}{|c|c|c|c|c|c|c|}
\hline
 & $|b_\ell|$ & $\angle b_\ell$ & $\phi_\ell$ & $\theta_\ell$ & $\omega_{1,\ell}$ & $\omega_{2,\ell}$\\
\hline
MSE & 5.5E-7 & 0.0241 & 0.0146 & 0.0191 & 4.5E-6 & 1.9E-4 \\
RMSE & 7.4E-4 & 0.1554 & 0.1210 & 0.1383 & 0.0021 & 0.0139 \\
\hline
\end{tabular}}
\label{tab:RMSE_ml}
\end{center}
\end{table}

\begin{figure}[!htbp]
\centering
\begin{subfigure}{.5\linewidth}
  \centering
  \includegraphics[width=0.7\textwidth]{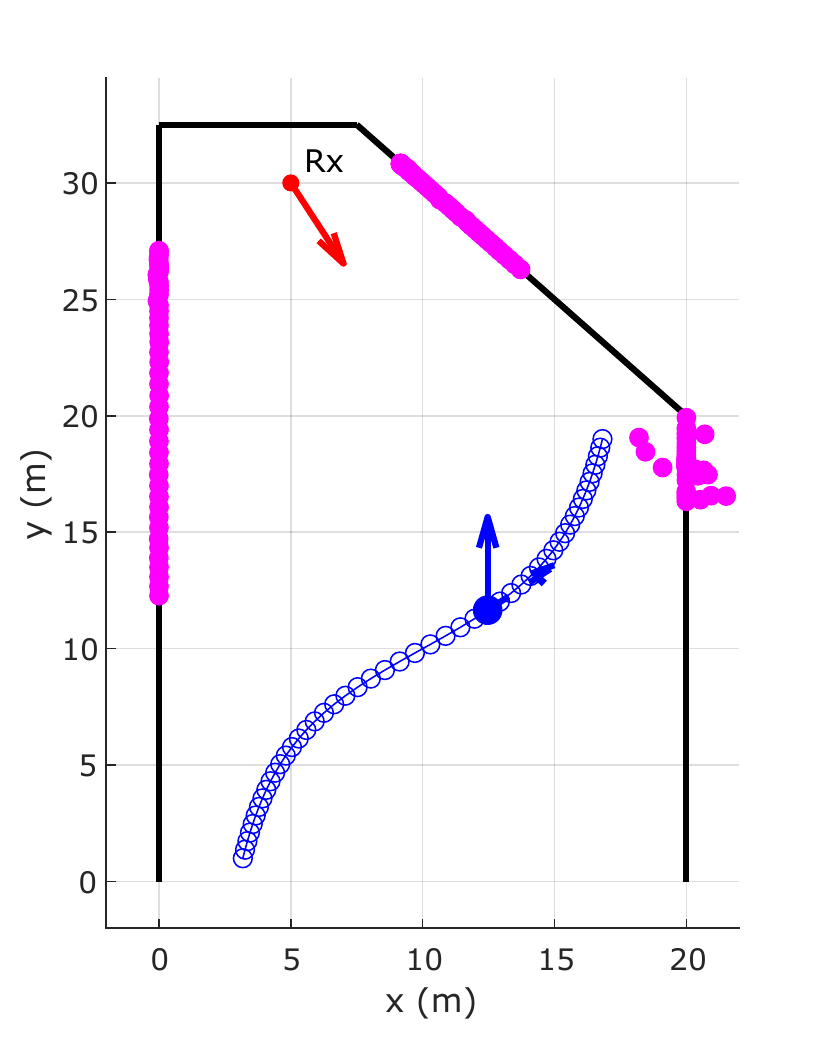}
  \caption{MAP}
  \label{fig:sub1_seq}
\end{subfigure}%
\begin{subfigure}{.5\linewidth}
  \centering
  \includegraphics[width=0.7\textwidth]{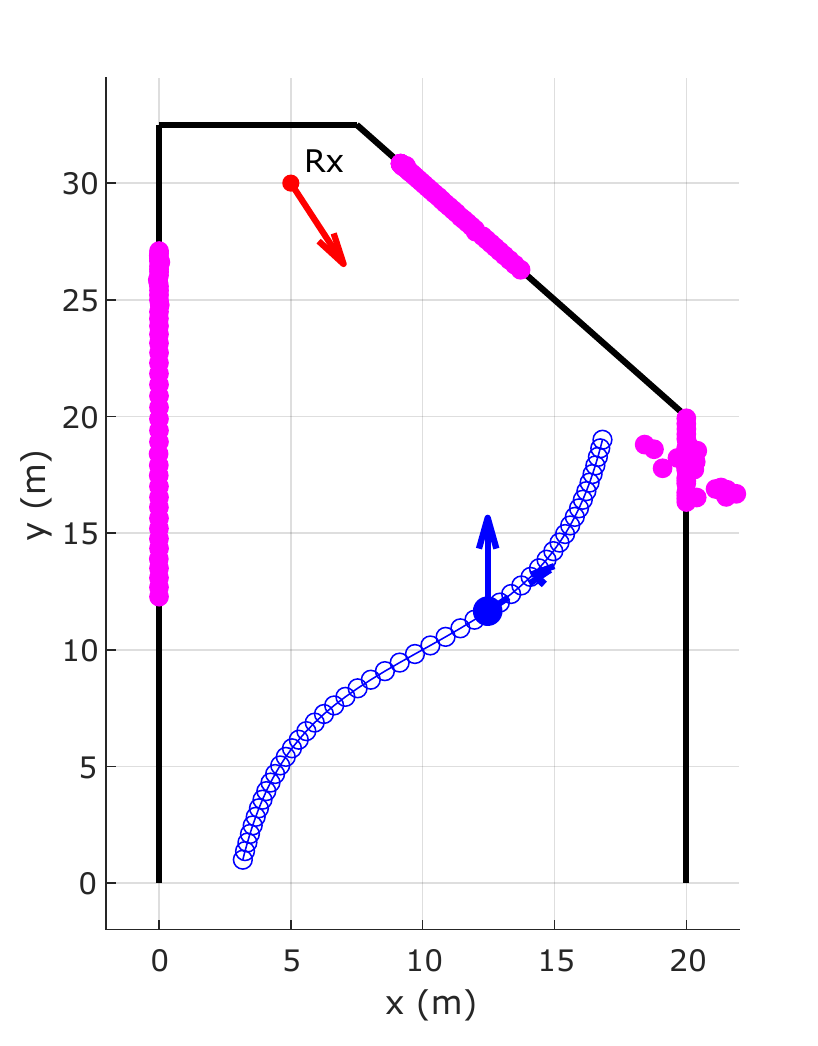}
  \caption{ML}
  \label{fig:sub2_seq}
\end{subfigure}
\caption{Environment and transmitter trajectory, \ac{UE} orientation is always facing north, i.e., $90^\circ$, \ac{BS} orientation is the same as in the first simulation. An example sample point is shown.}
\label{fig:sequential_est}
\end{figure}

Finally, we explore the sequential estimation scenario, in which the estimates from the previous instant are used as priors for the next estimation round. The path of the transmitter and the estimated position of reflectors using the line intersection method is shown in Fig.~\ref{fig:sequential_est}. The whole path is traveled over 5 seconds with 50 estimation rounds performed in equal time intervals. We set all variance parameters to $\nu = 0.005$ and achieve 1 precision 0.9844 recall. The equivalent \ac{ML} precision and recall are 0.9844 and 0.9844, respectively. The \ac{MSE} and the \ac{RMSE} values for \ac{MAP} and \ac{ML} in this scenario are presented in Table \ref{tab:RMSE_MAP}. Besides the improved precision and recall and similar \ac{MSE} values, the \ac{MAP} also converges faster, which can be beneficial in real time applications. It is up to the user to decide the best approach for the intended use.

\vspace{-0.3cm}
\begin{table}[!h]
\caption{\ac{MSE} and \ac{RMSE} for the parameter estimates of valid paths, \ac{MAP} and \ac{ML} estimation.}
\begin{center}
\resizebox{\columnwidth}{!}{
\begin{tabular}{|c|c|c|c|c|c|c|}
\hline
 & $|b_\ell|$ & $\angle b_\ell$ & $\phi_\ell$ & $\theta_\ell$ & $\omega_{1,\ell}$ & $\omega_{2,\ell}$\\
\hline
MSE (MAP)& 1.2E-6 & 0.1023 & 0.0093 & 0.0048 & 1.8E-5 & 1.0E-3 \\
RMSE (MAP)& 0.0011 & 0.3198  & 0.0964 & 0.0691 & 0.0043 & 0.0319 \\
MSE (ML) & 1.2E-6 & 0.0837 & 0.0093 & 0.0050 & 1.8E-5 & 6.2E-4 \\
RMSE (ML)& 0.0011 & 0.2894 & 0.0964 & 0.0708 & 0.0043 & 0.0250 \\
\hline
\end{tabular}}
\label{tab:RMSE_MAP}
\end{center}
\end{table}

\vspace{0.1cm}
\section{Conclusion}\label{sec:conclusion}
Estimating all the multipath components and their parameters is not a simple problem, and existing methods frequently rely on many simplifications or extensive computation that hinders its real-time applicability. In this paper, we have introduced a \ac{ML} and \ac{MAP} estimation procedure for channel estimation with possible use cases in sensing and mapping using an \ac{OFDM} waveform. The proposed method specifically exploits the problem structure and can be improved in straightforward fashion to provide increased robustness, efficiency, accuracy and detection capabilities. 
\section*{Acknowledgements}
The work was supported in part by the Research Council of Finland (former Academy of Finland) 6G Flagship Program (Grant Number: 346208) and 6GWiCE project (357719). We would also like to thank Hamza Djelouat, Mikko Sillanpää, and Reijo Leinonen for the productive discussions.

\ifCLASSOPTIONcaptionsoff
  \newpage
\fi



%

\bibliographystyle{IEEEtran}
\bibliography{IEEEabrv,biblio}



%







\end{document}